\documentclass{article}
\usepackage[draft]{hyperref} 
\usepackage{setspace}
\usepackage{amsmath}
\usepackage{amssymb}
\usepackage{amsthm}
\usepackage{multirow}
\usepackage{amsfonts}
\usepackage{color}
\usepackage{graphicx}
\usepackage{subfigure}  
\usepackage[]{algorithmicx}
\usepackage{algpseudocode,algorithm}
\usepackage{cite}
\usepackage{mathtools}
\usepackage{stmaryrd}
\usepackage[mathscr]{euscript}
\usepackage{mathrsfs}

\usepackage{soul}

\usepackage[normalem]{ulem}
\usepackage{spconf,amsmath,graphicx}
\usepackage{mathtools}

\title{Adaptive Time-Channel Beamforming \\ for Time-of-Flight Correction}
\name{Avner~Shultzman, Oded Drori and Yonina C. Eldar
\thanks{This project has received funding from the European Research Council (ERC) under the European Union’s Horizon 2020 research and innovation programme (grant agreement No. 101000967) and from the Manya Igel Centre for Biomedical Engineering and Signal Processing.}
}
\address{Faculty of Math and Computer Science, Weizmann Institute of Science, Israel\\
Emails: \{avner.shultzman, oded.drori, yonina.eldar\}@weizmann.ac.il}
\begin{document}
%
\maketitle
\begin{abstract}
Adaptive beamforming can lead to substantial improvement in resolution and contrast of ultrasound images over standard delay and sum beamforming. Here we introduce the adaptive time-channel (ATC) beamformer, a data-driven approach that combines spatial and temporal information simultaneously, thus generalizing minimum variance beamformers. Moreover, we broaden the concept of apodization to the temporal dimension. Our approach reduces noises by allowing for the weights to adapt in both the temporal and spatial dimensions, thereby reducing artifacts caused by the media's inhomogeneities. We apply our method to in-silico data and show 12\% resolution enhancement along with 2-fold contrast improvement, and significant noise reduction with respect to delay and sum and minimum variance beamformers.
\end{abstract}
\section{Introduction}
\label{sec:intro}
Due to its non-invasive and non-radiating nature, along with its low cost, ultrasound (US) imaging is widely used in medical applications. 
In brightness-mode (B-mode) US imaging, an image is generated by transmitting a series of acoustic pulses from an array of transducer elements. The transmitted pulses propagate through different tissues, leading to a sequence of reflections and refractions which create echoes that are then detected by the same array. After acquiring the US signal, beamforming algorithms are used to generate a US image, by applying corresponding time delays and averaging them with tailored weights \cite{VanVeen1988Beamformig,Steinberg1992,Thomenius1996}. The process of applying the weights is referred to as apodization, and it plays a crucial role in determining the resolution and contrast of the output image. 

The most common beamforming algorithm is the delay-and-sum (DAS) beamformer, which applies a constant apodization in the spatial domain. Recently, several adaptive beamforming approaches have shown substantial improvement in contrast and resolution \cite{Viola2005,Synnevag2007,Holfort2007,Vignon2008,Synnevag2009,Yonina2019iMap}, over standard DAS. Contrary to DAS, the weights of adaptive beamformers during the apodization process, are not predetermined, but instead are extracted from the signal.

Many adaptive methods are based on the minimum-variance (MV) beamformer, devised by Capon \cite{Capon69}. This algorithm minimizes the variance of the received signal by attributing specific weights to each of the channels; i.e. the elements of the transducer array. Several improvements were proposed, including the eigenspace-based minimum variance (EBMV) beamformer \cite{Aliabadi2016EBMV}, the iMap beamformer \cite{Yonina2019iMap}, and the robust Capon beamformer \cite{Wang2005} amongst others \cite{Ferguson1998MVDR,Nilsen2009}. However, all those methods weigh neighboring time samples in a predetermined way, thus weakening the ability to reduce noises over the temporal axis.

This paper aims to generalize the MV beamformer by introducing the adaptive time-channel (ATC) beamformer, which weighs the channels and the neighboring time samples simultaneously. We propose a data-driven approach, based on the conventional MV beamformer, which also deals with noise distributed temporally, and generalizes the concept of apodization to the temporal dimension. The primary source of noise arises from the media's inhomogeneities, which lead to inaccurate time-of-flight (TOF) calculation during the beamforming process. We exemplify our method on in-silico data and compare it to DAS and MV beamformers, in term of contrast and resolution. We show 12\% resolution's enhancement along with doubled contrast.

The rest of the paper is organized as follows. Section \ref{sec:ATC} derives the algorithm for the ATC beamformer and discusses some of its fundamental properties. Section \ref{sec:results} shows how the ATC beamformer performs on in-silico data, and compares our method to DAS and MV beamformers in terms of resolution and contrast. Finally, we highlight the main conclusions in Section \ref{sec:conc}.

\section{ATC Beamformer}
\label{sec:ATC}
In this section, we first present the conventional MV beamformer. We then introduce the ATC beamformer, discuss its fundamental properties and present how it generalizes current adaptive approaches. Throughout the paper, we will use the $[\cdot]$ operator to denote the position in a vector or a matrix.

\subsection{MV beamformer}
Consider an array of $M$ equally spaced transducer elements. During the scanning process, the transducer array samples the acoustic wave at $T$ equally spaced time samples. Corresponding time delays are applied to the acquired data to obtain the TOF aligned signal.

Each temporal value in the TOF aligned signal, is equivalent to a single location in the US image. The beamformer computes the apodization for a location, and repeats this process to form the entire image. We will focus on a single location, and refer to it as the central location, or equivalently, the central time sample. To compute the variance of the signal, the beamformer considers $K$ neighboring time samples in each direction around the central time sample. Therefore, the total number of time samples used for the reconstruction is $K_{tot} = 2K + 1$. We denote the TOF aligned signal used in the beamforming by $\Phi \in \mathbb{R}^{M \times K_{tot}}$, where $\Phi[m,k]$ is the value at channel $m$ and offset $k$ from the central time sample ($k = 0$ represents the central time sample).

The MV beamformer aims to minimize the variance of the output image, by attributing data-dependent weights to the channels. The output of the MV beamformer is
\begin{equation}
    z_{MV} = \sum_{m=0}^{M-1} \mathbf{w}_{MV}[m] \Phi[m, k = 0]
    \label{eq:z_MV}
\end{equation}
where $z \in \mathbb{R}$ and $w_{MV} \in \mathbb{R}^{M \times 1}$ are the data-dependent weights. To compute the variance of the TOF aligned signal we first define the MV covariance matrix \cite{Synnevag2009}
\begin{equation}
    R_{MV} = \frac{1}{K_{tot}} \sum_{k=-K}^K \mathbf{\Phi}_k \mathbf{\Phi}_k^T
    \label{eq:R_MV}
\end{equation}
where $\mathbf{\Phi}_i \in \mathbb{R}^{M \times 1}$ is the time sample at offset $i$ of the TOF aligned signal. The MV weights are then the solution of the following minimization problem
\begin{equation}
    \min_{\mathbf{w}_{MV}} \ \ \mathbf{w}_{MV}^T R_{MV} \mathbf{w}_{MV}
    \label{eq:min_problem_MV}
\end{equation}
\begin{equation}
    \ \ \ \ s.t. \ \ \ \mathbf{w}_{MV}^T \mathbf{a} = 1 \ \ \ \ \ \ \ \ \ 
    \label{eq:constraint_MV}
\end{equation}
where $\mathbf{a} \in \mathbb{R}^{M \times 1}$ is a vector of ones.
The analytical solution of the above problem is \cite{Capon69}
\begin{equation}
    \mathbf{w}_{MV, opt} = \frac{R_{MV}^{-1} \mathbf{a}}{\mathbf{a}^T R_{MV}^{-1} \mathbf{a}}
    \label{eq:solution_MV}
\end{equation}

\subsection{ATC Analytical Derivation}
We aim to generalize the current MV approach, by attributing tailored weights to the entire TOF aligned signal, which contains $K_{tot}$ time samples, and not just the central time sample. As a result, our method compensates for inaccurate TOF calculations, at the cost of increasing the number of degrees-of-freedom by a factor of $K_{tot}$.

We define the output of the ATC beamformer at the central location in the image, $z$, to be
\begin{equation}
    z_{ATC} = \sum_{i=-K}^{K} \mathbf{\Phi}_i^T \mathbf{w}_i
    \label{eq:z}
\end{equation}
where $\mathbf{w}_i \in \mathbb{R}^{M \times 1}$ are the weights corresponding to $\mathbf{\Phi}_i$.

Based on the MV approach, we propose to solve the following minimization problem:
\begin{equation}
    \min_{\{\mathbf{w}_i\}_{i = -K} ^ K} \mathcal{L} = \sum_{i = -K}^K \sum_{j = -K}^K \mathbf{w}_i^T \mathbf{\Phi}_i A_{ij} \mathbf{\Phi}_j^T \mathbf{w}_j
    \label{eq:L_ATC}
\end{equation}
\begin{equation}
    \ \ \ \ s.t. \ \ \ \sum_{i = -K}^K \mathbf{w}^T_i \mathbf{a} = 1, \ \ \ \ \ \ \ \ \ 
    \label{eq:constraint}
\end{equation}
where $A \in \mathbb{R}^{K_{tot} \times K_{tot}}$ are temporal apodizations. Note that if we choose $A$ to be a matrix of ones the objective reduces to $\mathcal{L} = |z_{ATC}|^2$. Generalizing the apodization process also to the temporal dimension, revises current MV approaches, and captures a vaster range of physical phenomena.

The above problem can be reformulated as an MV minimization problem. To this end, we define $R_{ATC} \in \mathbb{R} ^ {M \cdot K_{tot} \times M \cdot K_{tot}}$ to be a block matrix of $K_{tot} \times K_{tot}$ blocks of size $M \times M$ each, such that the $(i,j)$ block is
\begin{equation}
    R_{ATC, ij} = A_{ij} \mathbf{\Phi}_i \mathbf{\Phi}_j^T.
    \label{eq:R_ATC_Block}
\end{equation}
The objective of the ATC minimization problem can then be written as
\begin{equation}
    \mathcal{L} = \mathbf{w}_{vec}^T R_{ATC} \mathbf{w}_{vec}
    \label{eq:L_ATC_matrix}
\end{equation}
where $\mathbf{w}_{vec} \in \mathbb{R} ^ {M \cdot K_{tot} \times 1}$ is a concatenation of the weight vectors $\mathbf{w}_i$ for $i \in [-K,K]$.

Similar to MV, the analytical solution is
\begin{equation}
    \mathbf{w}_{vec, opt} = \frac{R_{ATC}^{-1} \mathbf{a}_{ATC}}{\mathbf{a}_{ATC}^T R_{ATC}^{-1} \mathbf{a}_{ATC}}
    \label{eq:solution}
\end{equation}
where, $\mathbf{a}_{ATC} \in \mathbb{R}^{M \cdot K_{tot} \times 1}$ is a vector of ones. The optimal weights $\mathbf{w}_{i,opt} \in \mathbb{R}^{M \times 1}$ for $i \in [-K,K]$ can be extracted from $\mathbf{w}_{vec, opt}$.

To increase the robustness of the algorithm, we load the diagonal of $R$, with a factor proportional to its trace, as suggested in \cite{Synnevag2009}:
\begin{equation}
    R_{DL} = R_{ATC} + \epsilon \cdot tr(R_{ATC}) \cdot \mathbf{I}
    \label{eq:new_R}
\end{equation}
where $\epsilon \in \mathbb{R}^+$ is a small variable, and $\mathbf{I}$ is the $M \cdot K_{tot} \times M \cdot K_{tot}$ identity matrix.

The process of extracting the ATC weights and generating a single point in the image is summarized in  Algorithm \ref{alg:algorithm}. To construct the entire image, we apply Algorithm \ref{alg:algorithm} on each point of the image separately.
\begin{algorithm}[h]
\begin{algorithmic}
\caption{ ATC beamforming.}
\label{alg:algorithm}
\Statex \textbf{Input:} $\mathbf{\Phi}_i, \forall i \in [-K,K], A$ - The TOF aligned US signal and the temporal apodizations.
\State Compute $R_{ATC}$ from (\ref{eq:R_ATC_Block})
\State Diagonal loading $R_{DL} = R_{ATC} + \epsilon \cdot tr(R_{ATC}) \cdot \mathbf{I}$
\State Compute the solution $\mathbf{w}_{vec, opt} = \frac{R_{DL}^{-1} \mathbf{a}_{ATC}}{\mathbf{a}_{ATC}^T R_{DL}^{-1} \mathbf{a}_{ATC}}$.
\State Extract the optimal weights $\mathbf{w}_{i, opt}$
\State Apply the apodization $z_{ATC} = \sum_{i=-K}^{K} \mathbf{\Phi}_i^T \mathbf{w}_{i, opt} $
\Statex \textbf{Output:} $z_{ATC}$.
\end{algorithmic}
\end{algorithm}

In terms of computational complexity, our method inverts a $M \cdot K_{tot} \times M \cdot K_{tot}$ matrix, instead of inverting a $M \times M$ matrix, thus increasing the complexity by a factor of $K_{tot}^3$. However, $K_{tot}$ is a constant number, and consequently MV and ATC have a similar complexity.

\section{Results and Discussion}
\label{sec:results}
We evaluated the performance of the ATC beamformer on in-silico data. The simulated data consists of point reflectors, and was generated using Verasonics Vantage 256, L11-5v, with 128-elements transducer. Our method is compared to the DAS and MV beamformers. The same post-processing (gamma correction and log compression) was applied to all results. With respect to the specific implementation of the MV and ATC algorithms, we set the number of neighboring time samples to be $K_{tot} = 11$, and $\epsilon = 10^{-10}$. Also, we used a triangular temporal apodization:
\begin{equation}
    A_{ij} = K + 1 - max(|i|, |j|), \ \ \ \ \forall i,j \in [-K, K].
    \label{eq:A_used}
\end{equation}

\begin{figure}[t!]
    \centering
    \includegraphics[width=0.4\textwidth]{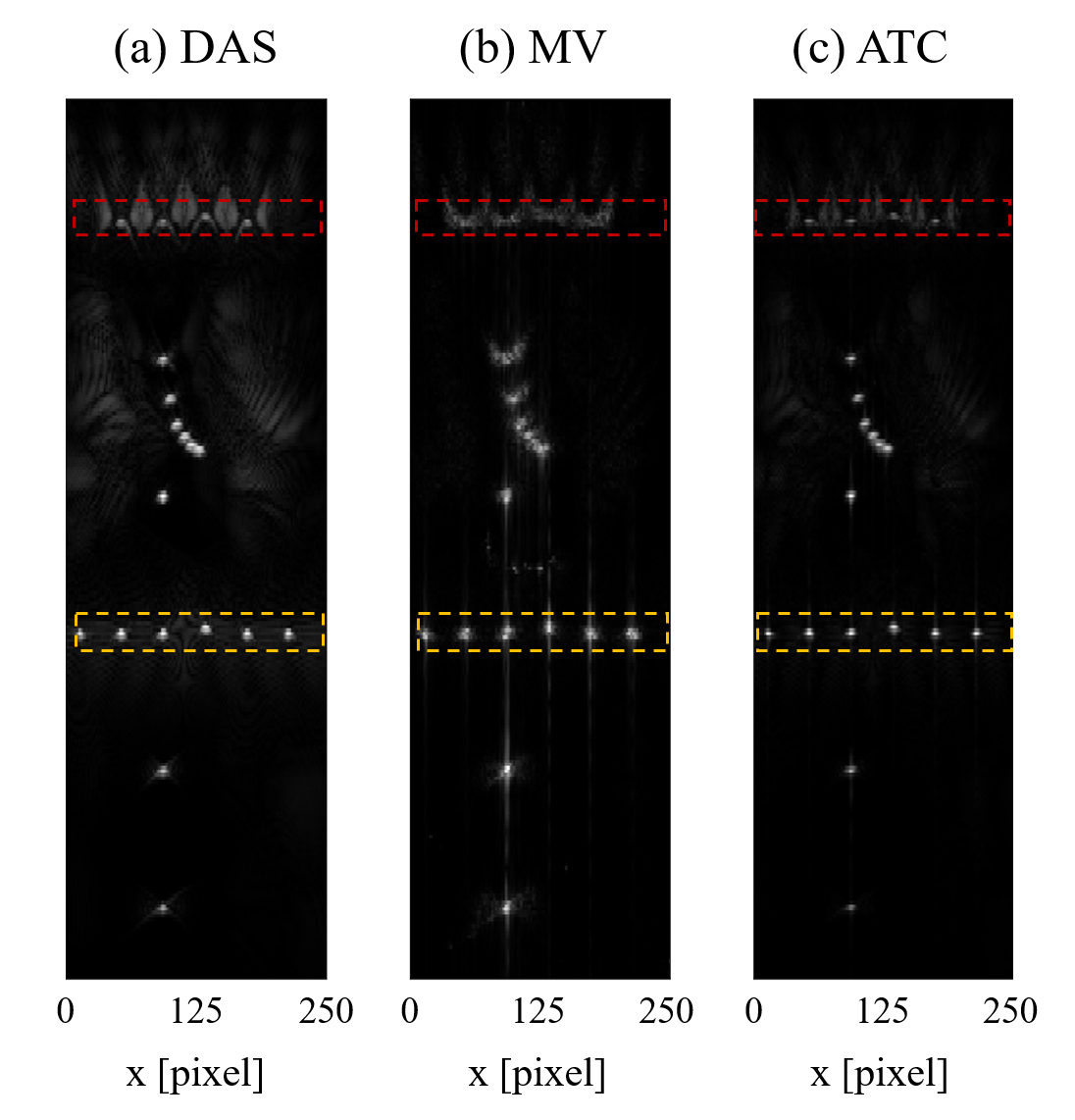}
    \caption{Beamformers comparison on in-silico data for $K = 5$. \textbf{(a)} DAS beamformer. \textbf{(b)} MV beamformer. \textbf{(c)} ATC beamformer. The transducer array is located at the top of the images. We notice high artifacts' intensity in the DAS reconstruction. The MV bemformer reduces those artifacts at the expense of lowering the resolution of the reflectors. The ATC beamformer overcomes this limitation by achieving noise reduction along with resolution enhancement.}
    \label{fig:Phantom}
\end{figure}

In Fig. \ref{fig:Phantom}, we compare the ATC beamformer to the DAS and MV beamformers. The image reconstructed using DAS suffers from strong artifacts at multiple locations in the image. The MV beamformer reduces those artifacts at the expense of lowering the resolution of the reflectors. Our approaches achieves artifacts' reduction along with resolution enhancement, thus improving over DAS and MV. 

To quantitatively compare the results we extract the point-spread-functions (PSF) of a chosen reflector, and calculate the resolution and contrast based on it. The resolution is defined as $1/$FWHM (full-width half-maximum), and the contrast is defined as the ratio between the primary and secondary lobes of the PSF. 
These quantitative measures depend on the choice of the reflector, and may vary within a single image.

\begin{figure}[t!]
    \centering
    \includegraphics[width=0.4\textwidth]{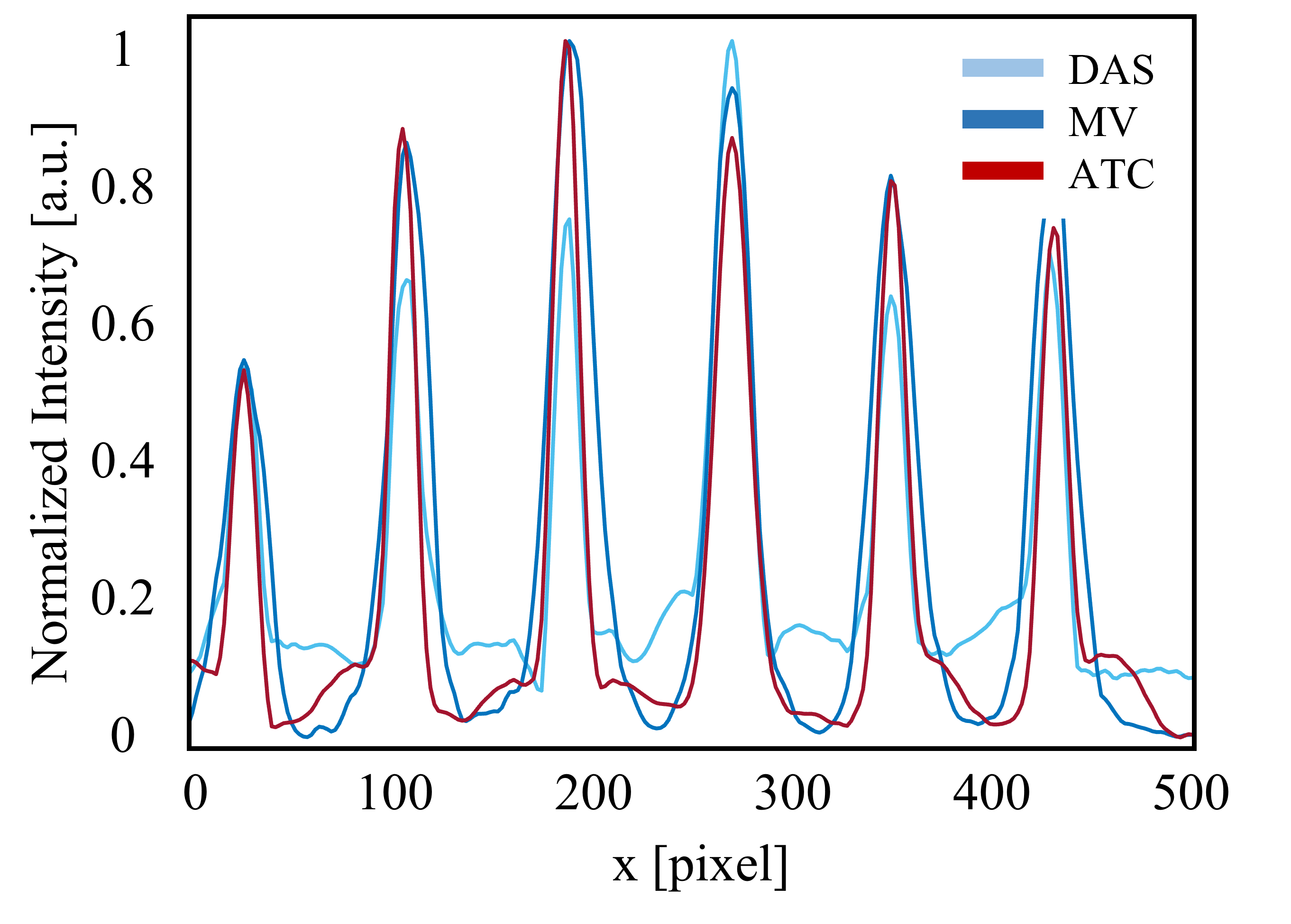}
    \caption{PSFs comparison - PSFs of the reflectors in the yellow box in Fig. \ref{fig:Phantom}, normalized by their maximal values. The resolution and contrast can be extracted based on the PSF. The ATC beamformer outperforms DAS and MV.}
    \label{fig:PSF}
\end{figure}

To demonstrate how the ATC beamformer improves the image, we present in Fig. \ref{fig:PSF}, the PSFs of the reflectors marked by a yellow box in Fig. \ref{fig:Phantom}, and consider the reflector at pixel 50. The  ATC beamformer improves its resolution by 12\% compared to DAS. Similarly, we consider the reflector at pixel 100, and compute its contrast. The ATC beamformer improves its contrast by a factor of 2 compared to DAS and 1.5 compared to MV. The results are summarized in Table 1. Moreover, the signal coming from the reflectors which are close to the transducer array (marked by a red box), is attenuated using the MV approach. Their visibility is restored with the ATC method, where temporal data is combined in the beamforming process.

\begin{table}[h!]
\centering
\begin{tabular}{ |p{1.7cm}||p{1.7cm}|p{1.7cm}|p{1.7cm}|  }
     \hline
     Method& DAS & MV & ATC \\
     \hline
     Resolution & $\times$1 & $\times$0.84 & \textbf{$\times$1.12}\\
     Contrast & $\times$1 & $\times$1.57 & \textbf{$\times$2.14}\\
     \hline
\end{tabular}
\label{table:res}
\caption{Comparison between the DAS, MV and ATC beamformers  in terms of resolution and contrast, on the reflectors in Fig. \ref{fig:Phantom}. The values are normalized w.r.t. DAS.}
\end{table}

\begin{figure}[t!]
    \centering
    \includegraphics[width=0.4\textwidth]{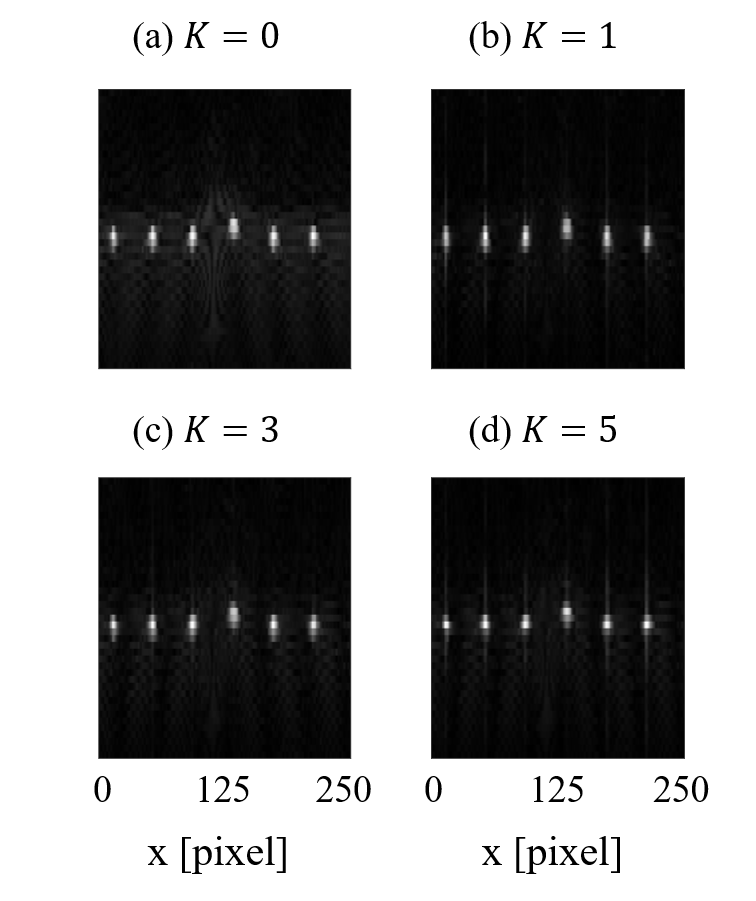}
    \caption{Dependence of the ATC beamformer on $K$. \textbf{(a)} $K = 0$. \textbf{(b)} $K = 1$. \textbf{(c)} $K = 3$. \textbf{(d)} $K = 5$. We notice that the intensity of the artifacts between the reflectors is reduced when additional time samples are considered ($K > 0$).}
    \label{fig:k_comp}
\end{figure}

Finally, in Fig. \ref{fig:k_comp} we show the dependence of the ATC reconstruction on the number of neighboring time samples taken into account, $K$. A significant noise reduction is observed when temporal information is combined in the beamforming process ($K > 0$). We computed the resolution of the reflectors in the image, and showed that resolution improves as we increase $K$. As we can see, the choice of $K$ plays an important role in the reconstruction process, and affects the resolution and contrast of the formed image.

\begin{table}[t!]
\centering
\begin{tabular}{ |p{1.4cm}||p{1.24cm}|p{1.24cm}|p{1.24cm}|p{1.24cm}|  }
     \hline
     Reflector location [pixel]& $K = 0$ & $K = 1$ & $K = 3$ & $K = 5$ \\
     \hline
     100 & $\times$1 & $\times$1.09 & $\times$1.20 & \textbf{$\times$1.39}\\
     220 & $\times$1 & $\times$1.12 & $\times$1.19 & \textbf{$\times$1.47}\\
     \hline
\end{tabular}
\label{table:res_k_comp}
\caption{Comparing the resolution of the reflectors in Fig. \ref{fig:k_comp} for different values of $K$. The results are normalized w.r.t. $K = 0$.}
\end{table}

\section{Conclusion}
\label{sec:conc}
We presented a data-driven approach for adaptive beamforming that combines spatial and temporal information simultaneously. Using the temporal information, we reduce artifacts induced by inhomogeneities in the media, and compensate for inaccurate TOF calculation. We demonstrated our approach on in-silico data and showed 12\% resolution enhancement along with a 2-fold contrast improvement, compared to the conventional MV beamformer.

\section{Compliance with Ethical Standards}
This is a numerical simulation research for which no ethical approval was required.

\bibliographystyle{IEEEbib}
\bibliography{paper}

\end{document}